\documentclass[a4paper,twocolumn,11pt]{quantumarticle}
\pdfoutput=1
\usepackage[utf8]{inputenc}
\usepackage[english]{babel}
\usepackage[T1]{fontenc}
\usepackage{amsmath}
\usepackage{hyperref}
\usepackage{tikz}
\usepackage{lipsum}
\usepackage{graphicx} % Explicitly included for image handling

\begin{document}

\title{Hybrid Quantum-Classical Optimisation of Traveling Salesperson Problem}

\author{Christos Lytrosyngounis}
\affiliation{Qosmos Technologies Research Center, 18 Irodotou Street, Athens, Attica 10675, Greece}
\orcid{0009-0006-0368-3494}
\author{Ioannis Lytrosyngounis}
\affiliation{Qosmos Technologies Research Center, 18 Irodotou Street, Athens, Attica 10675, Greece}

\maketitle

\begin{abstract}
The \textbf{Traveling Salesperson Problem (TSP)} is a fundamental NP-hard optimisation challenge with widespread applications in logistics, operations research, and network design. While classical algorithms effectively solve small to medium-sized instances, they struggle with scalability due to exponential complexity. In this work, we present a \textbf{hybrid quantum-classical approach} that leverages IBM’s Qiskit Runtime to integrate quantum optimisation techniques with classical machine learning (ML) methods, specifically \textbf{K-Means clustering and Random Forest classifiers}. These ML components aid in problem decomposition and noise mitigation, improving the quality of quantum solutions.

Experimental results for \textbf{TSP instances ranging from 4 to 8 cities} reveal that the \textbf{quantum-only approach produces solutions that are up to 21.7\% worse than the classical baseline}, while the \textbf{hybrid method reduces this cost increase to 11.3\%} for 8 cities. This demonstrates that hybrid approaches improve solution quality compared to purely quantum methods but still remain \textbf{suboptimal compared to classical solvers}. Despite current hardware limitations, these results highlight the potential of quantum-enhanced methods for combinatorial optimisation, paving the way for future advancements in scalable quantum computing frameworks.
\end{abstract}

\section{Introduction}
\label{sec1}
The TSP is one of the most extensively studied combinatorial optimisation problems in operations research and computer science due to its theoretical significance and practical applications in logistics, network design, and operations planning. The objective is to determine the shortest possible route that visits a given set of cities exactly once and returns to the starting point. As an NP-hard problem \cite{vanLeeuwen1998}, the computational complexity of solving the TSP grows exponentially with the number of cities, presenting significant challenges for classical optimisation techniques as problem size increases \cite{hybridTSP2024}. While heuristic and exact solvers have been developed to address the TSP, their effectiveness diminishes for large-scale instances due to the exponential growth of the solution space and the associated computational resource requirements. Recent advancements in quantum computing have introduced new paradigms for addressing complex optimisation problems. Quantum Approximate Optimisation Algorithms (QAOA) and Variational Quantum Algorithms (VQAs) \cite{QAOA2014} are two promising quantum techniques that leverage quantum mechanics to explore solution spaces more efficiently than their classical counterparts. However, limitations in current quantum hardware, including noise, connectivity constraints, and gate fidelity, hinder their direct application to real-world problems. These limitations necessitate hybrid quantum-classical approaches, which combine the strengths of quantum computing for exploration with classical computing for refinement. This study investigates a hybrid quantum-classical workflow for solving the TSP, leveraging IBM’s Qiskit Runtime environment \cite{hybridOptimization2024, ibmEstimatorV2}, to execute quantum circuits on simulators and physical quantum processors. Additionally, ML techniques are incorporated to optimise algorithm parameters and enhance solution accuracy, addressing the variability and reliability issues inherent in current quantum systems. The primary contributions of this work include:

\begin{itemize}
    \item A comparative evaluation of classical and quantum approaches for solving TSP instances ranging from 4 to 8 cities.
    \item An analysis of solution cost variability and robustness across different methods.
    \item Demonstration of the performance improvements achieved by integrating ML into the hybrid quantum-classical workflow.
\end{itemize}

The results underscore the potential of hybrid quantum-classical methods to outperform classical solvers for specific problem sizes, while identifying current limitations in quantum computing and proposing areas for future research and development.

This paper is organised as follows: 
\begin{itemize}
    \item \textbf{Section 2} (Background and Related Work): Introduces the fundamental aspects of the TSP and reviews classical, quantum, and hybrid strategies for solving it. This section discusses key developments in quantum algorithms, noise mitigation techniques, and the integration of machine learning to optimise hybrid quantum-classical workflows.
    \item \textbf{Section 3} (Methodology): Details the hybrid quantum-classical framework employed in this study, including the problem setup, quantum and classical computational components, and the role of machine learning in optimising parameter selection. This section also explains transpilation techniques, execution strategies, and noise mitigation methods.
    \item \textbf{Section 4} (Results and Analysis): Analyses the effectiveness of the hybrid approach in terms of cost efficiency and runtime performance compared to classical methods. The section includes a scalability study that highlights the feasibility and limitations of quantum-enhanced solutions for increasing problem sizes.
    \item \textbf{Section 5} (Discussion and Limitations): Examines the broader implications of the findings, identifying key constraints and limitations of current quantum hardware, the role of machine learning in improving solution stability, and potential pathways for enhancing hybrid optimisation techniques. This section discusses the future evolution of hybrid quantum-classical frameworks.
\end{itemize}

\section{Background and Related Work}
\label{background}
The TSP is a cornerstone in combinatorial optimisation, recognised for its theoretical significance and practical applications in logistics, network design, and operations research. Classical solvers, including exact methods such as branch-and-bound and heuristics like simulated annealing and genetic algorithms, provide effective solutions for small to medium-sized instances. However, as the problem scales, these methods face significant computational challenges due to the exponential growth of the solution space, which demands vast computational resources. Recent advancements in quantum computing present a novel paradigm for addressing combinatorial optimisation challenges such as TSP \cite{quantumChallenges2024}. QAOA and other variational quantum algorithms leverage the principles of quantum mechanics to explore solution spaces more efficiently than classical methods. These quantum techniques, when integrated into hybrid quantum-classical frameworks, have shown great potential for solving problems that are computationally infeasible for classical solvers. Prior research highlights the advantages of hybrid quantum-classical workflows, where quantum circuits are iteratively optimised using classical algorithms \cite{QAOAML2024, quantumMLReview}. Such frameworks have demonstrated results competitive with traditional solvers for specific problem instances. A critical component of these workflows is circuit transpilation, which maps quantum circuits onto hardware-specific constraints, such as qubit connectivity and native gate sets, while minimising circuit depth and error rates. The choice of transpilation strategies and optimisation levels significantly influences the performance of quantum algorithms, especially on noisy intermediate-scale quantum (NISQ) devices. ML has further enhanced hybrid approaches by improving the efficiency and robustness of optimisation processes. For instance, ML techniques have been shown to fine-tune QAOA parameters, resulting in reduced variability and improved solution quality. Additionally, ML models can assist in identifying optimal transpilation strategies and predicting execution outcomes, thereby addressing hardware limitations. Building upon this foundation, the present study explores the application of IBM’s Qiskit Runtime primitives for solving TSP instances. By integrating advanced quantum primitives, such as SamplerV2 and EstimatorV2, with ML-driven parameter optimisation, the study evaluates the scalability and robustness of hybrid quantum-classical workflows.

\section{Methodology}
\subsection{Problem Setup}
\label{sec3}
This study explores a hybrid quantum-classical workflow to solve the TSP, targeting instances with 4 to 8 cities. This range was deliberately selected to balance the constraints of current NISQ hardware—such as the IBM Sherbrooke processor (127 qubits, Eagle r3 architecture)—with the need to gain insights into the performance and scalability of quantum-enhanced methods. Encoding an \( n \)-city TSP as a Quadratic Unconstrained Binary Optimisation (QUBO) problem demands approximately \( n^2 \) qubits for binary decision variables (\( x_{ij} \)), plus additional qubits for constraints. With noise accumulation, limited circuit depth, and decoherence further degrading solution quality beyond 8 cities, this scope ensures reliable experimentation while aligning with NISQ capabilities. Simultaneously, 4–8 cities span a combinatorially meaningful range enabling exhaustive classical benchmarking and revealing scalability trends for hybrid approaches. While 4–8 cities represent a narrow scope, they serve as a critical stepping stone for evaluating the hybrid approach’s potential generalisability.

On the order of \( n^2 \) qubits are required by the QUBO formulation for an \( n \)-city TSP. For 9 cities, 81 qubits are needed for the decision variables, along with additional \textit{ancilla} or \textit{penalty} qubits to encode constraints. Ancilla qubits are auxiliary qubits used to enforce constraints via intermediate computations, while penalty qubits introduce additional energy terms in the Hamiltonian to discourage invalid solutions. These additional qubits help ensure that each city is visited exactly once and that a valid TSP tour is formed, but they increase the overall quantum resource requirements. In practice, this pushes the requirements near or beyond what the 127-qubit IBM Sherbrooke processor can reliably handle. An attempt to solve a 9-city instance on a simulator was made, but it was found to fail due to resource constraints (the statevector simulation became intractable and ran out of memory). On real hardware, even if enough qubits are available, the execution of a QAOA circuit with approximately 80 or more qubits and the necessary two-qubit gates is hampered by unmanageable error rates—effectively preventing the circuit from running to completion before decoherence erases the quantum advantage.

\subsection{Classical TSP Solver}
\label{sec3_2}
The classical baseline solver used in this study is a \textbf{brute-force enumeration} of all possible TSP tours. The optimal solution is determined by iterating over all city permutations and computing the total travel cost for each possible route. The implementation follows:

\begin{equation}
    C(T) = \sum_{i=1}^{N-1} d(c_i, c_{i+1}) + d(c_N, c_1)
\end{equation}

where \( C(T) \) is the total cost of the tour, \( c_i \) represents a city in the sequence, and \( d(c_i, c_j) \) is the travel distance between two cities.

\begin{figure}[htbp]
    \centering
    \includegraphics[width=\columnwidth, keepaspectratio]{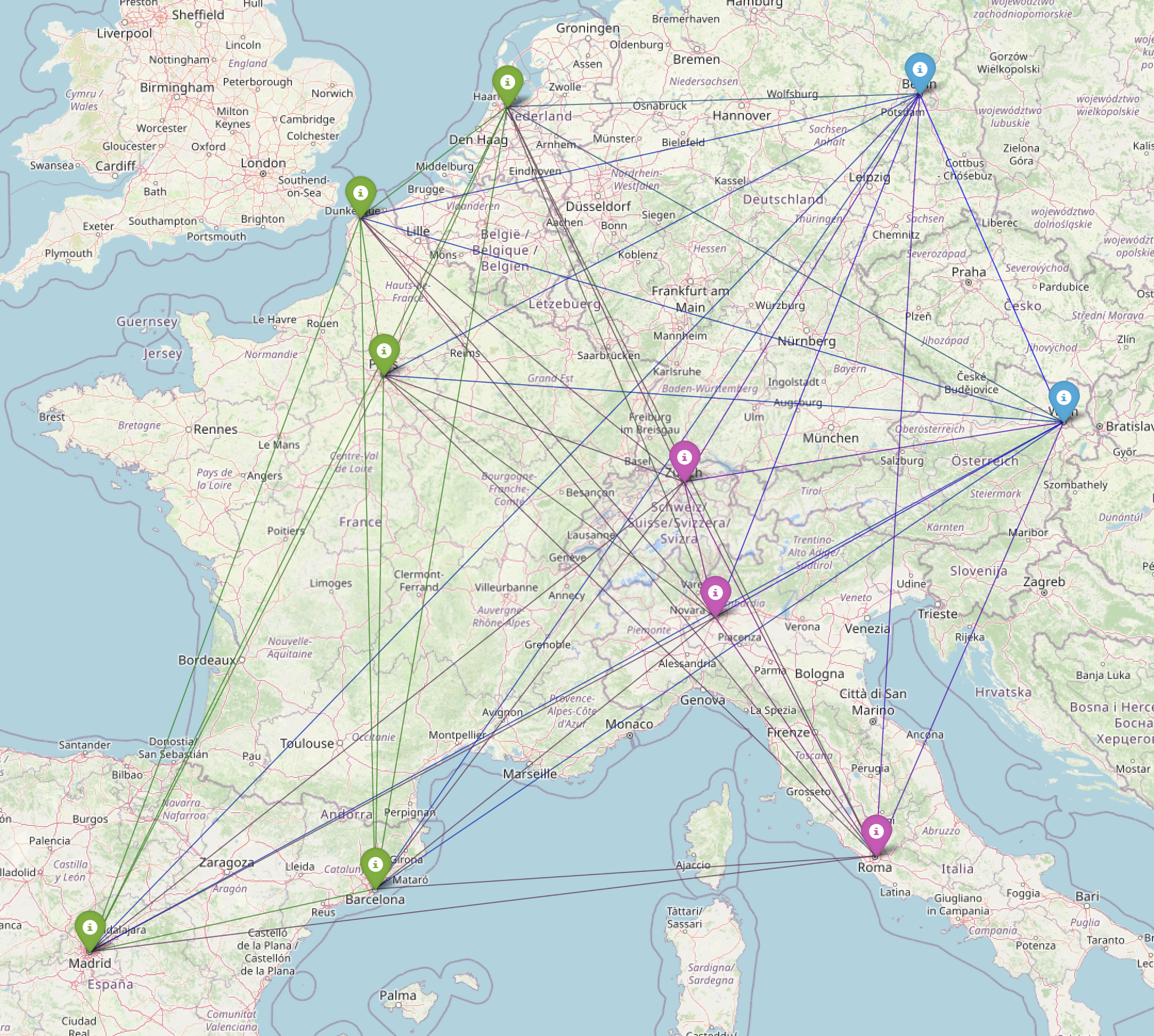}
    \vspace{-10pt}
    \caption{Graph representation of a TSP instance. Cities are depicted as nodes, while weighted edges represent pairwise travel distances. The objective is to determine the shortest possible route that visits each city exactly once before returning to the starting point. The fully connected structure reflects the assumption that direct travel is possible between any two cities.}
    \label{fig:1}
\end{figure}

Since the number of possible tours grows factorially (\( O(n!) \)), this method is computationally feasible only for small instances (\( n \leq 8 \)). For larger problems, heuristics or meta-heuristics (e.g., Lin-Kernighan, Genetic Algorithms) are commonly used to approximate near-optimal solutions efficiently. However, this work focuses on a hybrid quantum-classical approach rather than heuristic-based classical optimisation.

The TSP was modelled as a fully connected weighted graph, where cities were represented as nodes and travel distances were assigned as edge weights. As presented in Figure~\ref{fig:1}, the problem instance was based on a predefined set of \textbf{ten European cities}:

\begin{itemize}
    \item \textbf{Amsterdam} (4.9041, 52.3676)
    \item \textbf{Barcelona} (2.1734, 41.3851)
    \item \textbf{Berlin} (13.4050, 52.5200)
    \item \textbf{Calais} (1.8587, 50.9513)
    \item \textbf{Madrid} (-3.7038, 40.4168)
    \item \textbf{Milan} (9.1900, 45.4642)
    \item \textbf{Paris} (2.3522, 48.8566)
    \item \textbf{Rome} (12.4964, 41.9028)
    \item \textbf{Vienna} (16.3738, 48.2082)
    \item \textbf{Zurich} (8.5417, 47.3769)
\end{itemize}

To ensure consistency across all experiments, \textbf{Calais} was designated as the departure city and \textbf{Milan} as the destination city for all TSP instances. This selection not only ensures a uniform problem definition across classical, quantum, and hybrid methods but also serves to reduce the computational complexity of the experiment, facilitating execution within the constraints of NISQ hardware. The standard TSP requires evaluating all possible permutations of cities, a solution space of size \( (n-1)! \) for an \( n \)-city instance when the starting point is fixed, which translates to a quadratic increase in qubit requirements (\( n^2 \)) when encoded as a QUBO problem. By fixing both a departure (Calais) and a destination (Milan), we constrain the solution space, reducing the number of valid tours that must be explored. For instance, in an 8-city TSP, the unconstrained problem has \( (8-1)! = 5,040 \) possible tours from a fixed start, but fixing an intermediate destination imposes a partial order, lowering the effective search space. This reduction is particularly advantageous for quantum execution, where circuit depth, gate count, and qubit count directly influence error rates and decoherence. By decreasing the computational burden, this approach enables more reliable circuit execution within the coherence time limits of the Sherbrooke processor (e.g., \( T_1 \approx 262.75 \, \mu \)s, \( T_2 \approx 169.99 \, \mu \)s) and reduces the transpilation overhead. The classical refinement step also benefits from a smaller set of candidate solutions, enhancing overall efficiency. While this introduces a constrained TSP variant, the inclusion of the return leg from Milan to Calais ensures the problem remains a closed tour, aligning with TSP’s core objective while making the experiment more feasible for current quantum hardware. The number of cities per instance was varied between \textbf{4 and 8}, selected dynamically from this set.

\begin{figure}[htbp]
    \centering
    \includegraphics[width=\columnwidth, keepaspectratio]{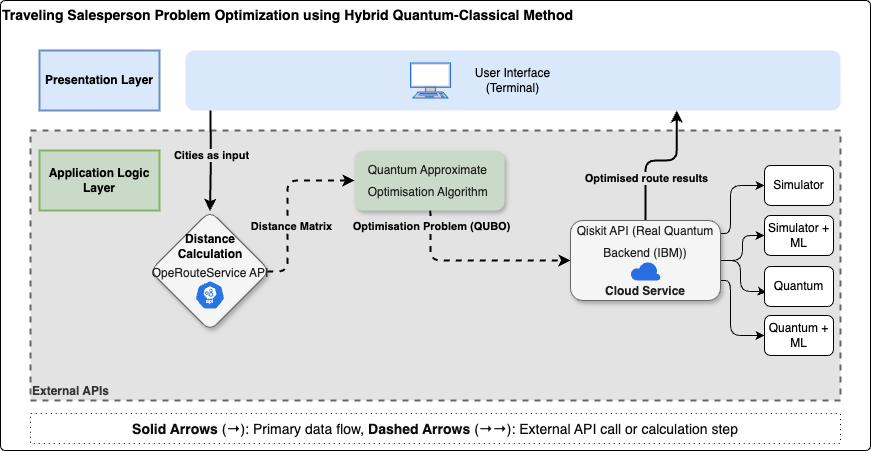}
    \vspace{-20pt}
    \caption{Traveling Salesperson Problem Optimisation - Hybrid Quantum-Classical Method Architecture. An overview of the hybrid quantum-classical workflow architecture used to solve the TSP. The diagram highlights the integration of quantum and classical components, showing how quantum circuits are used to explore solutions while classical algorithms refine and optimise them iteratively. The workflow includes transpilation, parameter optimisation, and noise mitigation.}
    \label{fig:2}
\end{figure}

To evaluate the efficacy of the hybrid quantum-classical approach, each TSP instance was solved using two distinct methodologies. First, a classical solver was employed as a baseline, using state-of-the-art heuristic algorithms to obtain solutions. Second, a quantum workflow was implemented, wherein the TSP instances were encoded into quantum circuits and executed on both quantum simulators and physical quantum hardware using IBM’s Qiskit Runtime environment. The Qiskit primitives, specifically SamplerV2 and EstimatorV2, were used to optimise quantum circuit executions under realistic constraints. The quantum workflow distributed optimisation tasks between quantum circuits and classical components, leveraging a hybrid approach. To enhance solution quality and address quantum hardware limitations, ML techniques were integrated into the workflow. ML models were used to optimise the variational parameters of the QAOA, improving the convergence rate and reducing variability in solution quality across executions. The ML-driven optimisation enabled more effective utilisation of the quantum hardware resources, particularly for NISQ devices. The experimental design systematically varied the number of cities within the range of 4 to 8, allowing a detailed analysis of the scaling behaviour and comparative performance of quantum-enhanced solutions against classical solvers. Performance metrics such as solution quality, cost variability, and scalability were used to evaluate the outcomes. Specifically, solution quality measured the total traversal cost, cost variability captured the consistency of results across multiple runs, and scalability assessed the performance as problem size increased. Figure~\ref{fig:2} provides an overview of the hybrid quantum-classical architecture implemented in this study.

Through this setup, the study analyses the strengths and limitations of quantum-enhanced methods, offering insights into their potential for solving combinatorial optimisation problems and their competitive advantage over classical approaches under specific conditions.

\subsubsection{Quantum Component}
The quantum workflow uses IBM’s Qiskit Runtime primitives, specifically SamplerV2 and EstimatorV2, to encode and execute the TSP problem. Parameterised quantum circuits are constructed using quantum gates, where potential solutions are represented and iteratively optimised. These circuits are executed on both quantum simulators and the IBM Sherbrooke quantum processor, a 127-qubit device built on the Eagle r3 architecture. Transpilation ensures compatibility with hardware-specific constraints, including qubit connectivity and gate sets, while minimising circuit depth to address limitations imposed by noise and decoherence. The IBM Sherbrooke backend was selected for its advanced capabilities, including high qubit count, competitive gate error rates, and robust connectivity. Single-qubit gates exhibit error rates of approximately \(2.726 \times 10^{-4}\), while two-qubit operations such as the ECR gate show median errors of \(7.984 \times 10^{-3}\). Relaxation (\(T_1\)) and dephasing (\(T_2\)) times, with medians of \(262.75 \, \mu s\) and \(169.99 \, \mu s\), constrain circuit execution time and depth, necessitating careful scheduling and optimisation.

\subsubsection{Classical Component}
The Classical Component serves as a crucial counterpart to the quantum component, playing a dual role in both evaluating candidate solutions and iteratively refining quantum circuit parameters. After the quantum execution stage, the sampled routes are classically assessed by computing their corresponding travel costs, ensuring feasibility and solution quality. This cost evaluation serves as feedback for updating the quantum optimisation parameters, enhancing solution convergence over multiple iterations. To further \textbf{improve efficiency and scalability}, two ML techniques—detailed in the following subsection—are integrated into the workflow. These ML methods help:
\begin{itemize}
    \item \textbf{Reduce the computational burden} by guiding the search towards high-quality solutions, thereby minimising unnecessary quantum circuit evaluations.
    \item \textbf{Optimise quantum resource utilisation} by mitigating the impact of noise and improving parameter selection in variational quantum algorithms.
\end{itemize}

By leveraging these \textbf{classical and ML-driven enhancements}, the \textbf{hybrid workflow} enables the effective resolution of \textbf{TSP instances within the constraints of NISQ hardware}, addressing current limitations in quantum hardware while improving solution reliability.

\subsection{Integration of Machine Learning Techniques}
The hybrid quantum-classical approach to the TSP distinguishes itself by integrating classical ML techniques—specifically, K-Means clustering and Random Forest classifiers—into the quantum optimisation process. This fusion enhances solution quality and bolsters resilience against quantum hardware noise.

\subsubsection{Distinctive Use of Machine Learning}
\begin{figure}[htbp]
    \centering
    \includegraphics[width=\columnwidth, keepaspectratio]{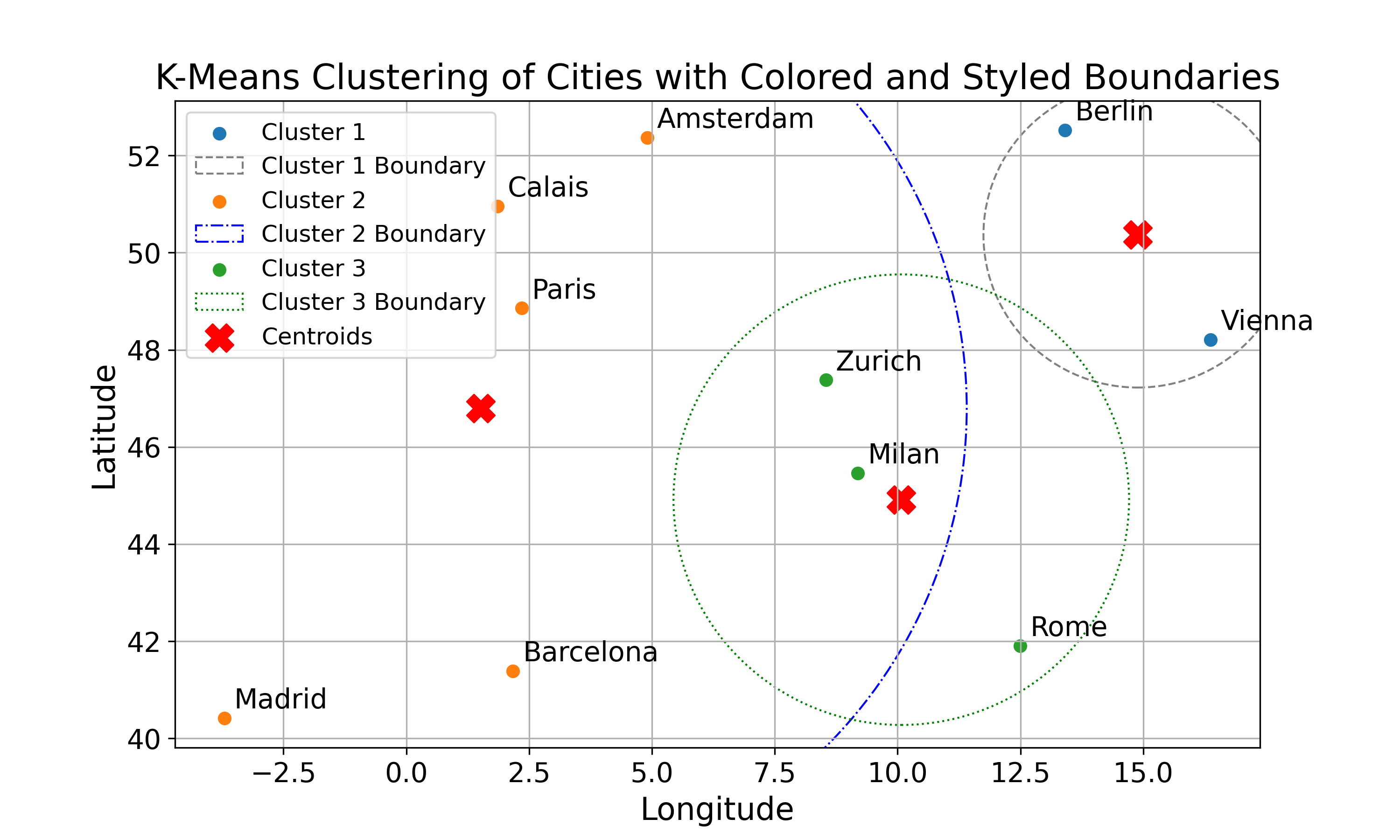}
    \vspace{-20pt}
    \caption{Visualization of K-Means clustering applied to the TSP. Cities are grouped into distinct clusters, represented with unique colours and styled boundaries. Red crosses mark the centroids of each cluster. This preprocessing step reduces the problem complexity, enabling efficient utilisation of quantum and classical resources for smaller subproblems.}
    \label{fig:3}
\end{figure}

\paragraph{K-Means Clustering for Problem Decomposition:} We employ K-Means clustering to partition cities into smaller, more manageable clusters, reducing the complexity of quantum circuits required. This decomposition enables efficient handling of subproblems suitable for current quantum hardware. Cities are grouped into smaller, geographically proximate subsets using K-Means clustering~\cite{kMeansModernPower}. The algorithm takes city coordinates (latitude, longitude) as input, with the number of clusters set to \( k = 3 \) (chosen heuristically to balance subproblem size and computational feasibility for 4–8 cities). The number of clusters was chosen to ensure subproblem sizes remain within feasible quantum limits. A larger \( k \) would reduce each cluster’s complexity but increase the complexity of merging sub-solutions. With 4–8 cities in total, setting \( k=3 \) means each cluster contains on average 2 or 3 cities, which yields subproblems small enough to be handled by the quantum algorithm (QAOA) on NISQ hardware. Using fewer clusters (e.g., \( k=2 \)) would result in larger clusters (up to 4 cities in one cluster when there are 8 total), potentially exceeding the practical quantum circuit size for our hardware. Conversely, using more clusters (e.g., \( k=4 \) or more) would make each quantum subproblem even smaller (maybe 1–2 cities, trivial to solve) but would increase the complexity of stitching the sub-routes together into a global tour and likely offer diminishing returns. Hyper-parameters include the Euclidean distance metric and 100 maximum iterations for convergence, implemented via scikit-learn. No separate validation set is used, as the clustering serves as a preprocessing step to reduce problem complexity, with TSP solutions integrated globally post-clustering (Figure~\ref{fig:3}).

Figure~\ref{fig:3} illustrates the application of K-Means clustering, where cities are grouped into three distinct clusters, each represented with unique colours and boundary styles. The centroids, marked with red crosses, serve as representative points for their respective clusters. This clustering approach significantly reduces the computational burden, enabling the hybrid quantum-classical workflow to efficiently solve larger instances of TSP by addressing smaller subproblems and integrating their solutions.

\paragraph{Random Forest Classifiers for Noise Mitigation:} Random Forest classifiers analyse quantum measurement outputs, distinguishing between noise-induced errors and probable optimal solutions. This post-processing step enhances the accuracy of solutions derived from the quantum processor. A Random Forest regressor~\cite{QAOAML2024} predicts TSP costs from quantum outputs to guide parameter optimisation. Training data consist of feature-label pairs derived from SamplerV2 results: features are measurement frequencies (counts) of bit-strings, and labels are corresponding TSP costs computed from the distance matrix. The model uses 100 decision trees (n\_estimators = 100), a maximum depth of 10, and mean squared error as the loss function, tuned via 5-fold cross-validation to minimise overfitting. Trained on 50 quantum runs per instance, the regressor predicts costs for new bit-strings, stabilising convergence by prioritising lower-cost solutions.

In contrast, prior hybrid quantum-classical TSP approaches~\cite{DWavesolvers} have primarily focused on integrating quantum algorithms with classical optimisation techniques without explicitly incorporating ML methods for clustering or noise mitigation. For instance, some studies have explored hybrid algorithms combining quantum annealing with classical optimisation but did not utilise ML for problem decomposition or error correction.

\subsubsection{Workflow Execution}
The execution of the hybrid workflow begins with parameter initialisation, either randomly or using ML-informed values. Parameterised quantum circuits are then executed to sample candidate solutions from the solution space. The sampled solutions are evaluated classically to compute their associated costs, and the results are used to refine the quantum circuit parameters iteratively. This feedback loop continues until convergence, which is determined by achieving a predefined cost threshold or reaching a set iteration limit. The continuous interaction between quantum sampling and classical evaluation ensures systematic refinement of the solutions, enabling high-quality results for TSP instances.

\subsubsection{Transpilation and Optimisation}
Transpilation of quantum circuits is essential for efficient execution on quantum hardware, ensuring compatibility with hardware constraints such as qubit connectivity and gate fidelity. Qiskit’s \texttt{generate\_preset\_pass\_manager} function was employed with optimisation level 1, which balances gate count reduction and compilation time. In Qiskit, optimisation levels range from 0 (no optimisation) to 3 (heavy optimisation). Level 1 applies a moderate set of optimisations that include gate cancellation, light re-scheduling, and qubit mapping, without spending excessive compilation time. Optimisation level 1 was selected to balance circuit quality and compilation time. At level 1, the transpiler is able to reduce the circuit depth and gate count (which is crucial for running on hardware with limited coherence time) and to route the circuit respecting the hardware’s qubit connectivity, while not introducing overly complex transformations that could themselves be time-consuming or error-prone.

The transpilation process includes:
\begin{itemize}
    \item \textbf{Circuit Depth Reduction:} Simplifying circuit structure to minimise gate operations.
    \item \textbf{Layout and Routing:} Mapping logical qubits to low-error physical qubits based on calibration data.
    \item \textbf{Gate Scheduling:} Aligning operations to reduce idle times and optimising the timing of gates to mitigate decoherence effects.
\end{itemize}

\subsubsection{Noise Mitigation Strategies}
The inherent noise in NISQ devices presents challenges for reliable quantum computations. To address these challenges, a series of noise mitigation strategies were applied:

\begin{itemize}
    \item \textbf{Noise Simulation and Modelling:} Qiskit’s \texttt{AerSimulator} was used to model realistic noise effects by integrating a noise model derived from a simulated backend. This allowed for pre-execution testing to assess circuit robustness under noisy conditions before deployment on real quantum hardware.
    \item \textbf{Backend-Aware Circuit Optimisation:} Transpilation techniques were applied using Qiskit’s \texttt{generate\_preset\_pass\_manager}, optimising the circuit depth, gate count, and qubit selection based on the target backend. This helped minimise the effects of gate errors and decoherence by prioritising low-error qubits and efficient gate arrangements.
    \item \textbf{Noise-Aware Execution on Real Quantum Hardware:} Noise characteristics of IBM’s Sherbrooke backend were retrieved using \texttt{backend.properties()}, providing real-time data on qubit coherence times (T1/T2), gate errors, and readout fidelity. This information guided execution decisions to improve overall stability.
    \item \textbf{Machine Learning for Noise-Resilient Solution Selection:} A Random Forest regressor was trained on past quantum output distributions to predict the most stable and cost-effective TSP paths. By leveraging classical ML models, the search space was refined to reduce the impact of noise-induced errors in quantum results.
\end{itemize}

\subsubsection{Backend Selection and Performance}
The IBM Sherbrooke backend, based on the Eagle r3 architecture, was chosen for its advanced specifications, which include:
\begin{itemize}
    \item \textbf{127 qubits} for high parallelism.
    \item \textbf{Native gate set} compatibility (ECR, ID, RZ, SX, X gates).
    \item \textbf{Performance metrics} such as a \textbf{CLOPS} rate of 30,000 and low error rates across single- and two-qubit operations.
\end{itemize}

Regular calibration, approximately every 30 minutes, ensured consistent hardware performance during the experiments.

\section{Results and Analysis}

\subsection{Comparison of Costs Across Methods}
\begin{figure}[htbp]
    \centering
    \includegraphics[width=\columnwidth, keepaspectratio]{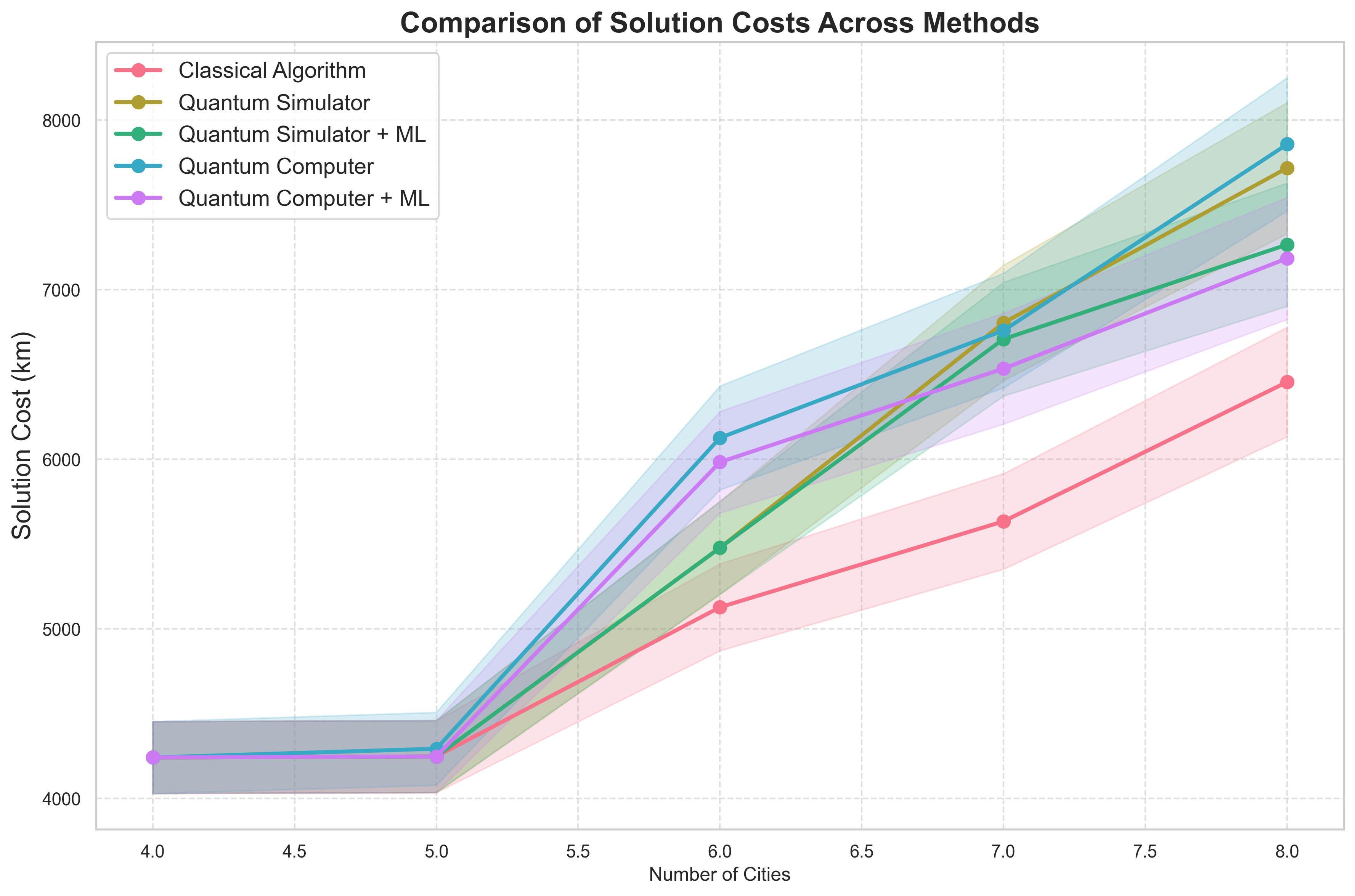}
    \vspace{-20pt}
    \caption{Solution cost comparison for classical, quantum, and hybrid quantum-classical methods for TSP instances of varying sizes (4–8 cities). The figure includes confidence intervals to show variability and highlights the competitive performance of hybrid approaches, especially for larger problem sizes.}
    \label{fig:4}
\end{figure}

The performance of classical solvers, quantum methods, and hybrid quantum-classical approaches was compared by analysing the solution costs for TSP instances with 4 to 8 cities. The comparison is summarised in Figure~\ref{fig:4}.

Classical methods consistently delivered near-optimal solutions for smaller problem instances. For TSP instances with 4–6 cities, classical solvers outperformed all other methods in terms of solution cost and consistency, as indicated by the steady and lower-cost trend. However, for larger problem sizes (7–8 cities), the performance of classical solvers began to degrade due to the increasing computational overhead required to explore the exponentially growing solution space. This limitation highlights the challenges of scaling classical methods for combinatorial optimisation problems. The standalone quantum approaches, including the quantum simulator and quantum computer (without ML enhancements), exhibited greater variability and higher solution costs compared to classical solvers. The variability, as evidenced by the broader uncertainty ranges, increased significantly for larger problem instances (6–8 cities). This behaviour is primarily attributed to the effects of hardware noise, limited qubit coherence, and suboptimal parameter tuning in current quantum algorithms. However, for 7–8 city instances, the gap between classical and quantum methods narrowed slightly, suggesting incremental improvements in quantum sampling, noise mitigation, and circuit optimisations. The integration of ML into the quantum workflow substantially improved solution quality and consistency. Both the quantum simulator + ML and quantum computer + ML methods demonstrated reduced solution variability and lower costs compared to their non-ML counterparts. The improvements were particularly pronounced for 6–8 city instances, where the hybrid approaches approached the performance of classical solvers. ML enhancements facilitated better parameter initialisation and convergence, helping to stabilise quantum outputs and mitigate the effects of noise.

\subsection{Performance Metrics and Results Analysis}
\label{sec:performance}

This section presents the performance evaluation of our hybrid quantum-classical approach for solving the TSP. We analyse solution cost variation, approximation accuracy, and scalability across different computational methods, including classical solvers, quantum-only approaches, and hybrid quantum-classical techniques.

\subsubsection{Performance Metrics}
To quantitatively assess the effectiveness of our approach, we define the following key performance indicators.

The cost variation percentage measures the expansion or reduction of the hybrid quantum-classical approach compared to classical solvers:

\begin{equation}
    \text{Cost Variation (\%)} = \frac{C_{\text{classical}} - C_{\text{hybrid}}}{C_{\text{classical}}} \times 100
\end{equation}

where:
\begin{itemize}
    \item \( C_{\text{classical}} \) is the cost (distance) of the optimal classical solution.
    \item \( C_{\text{hybrid}} \) is the cost obtained using the hybrid quantum-classical approach.
\end{itemize}

A lower percentage indicates better performance compared to classical methods.

The Approximation Ratio evaluates how close the quantum-enhanced solution is to the classical optimal solution:

\begin{equation}
    \text{Approximation Ratio} = \frac{C_{\text{quantum}}}{C_{\text{classical}}}
\end{equation}

where:
\begin{itemize}
    \item \( C_{\text{quantum}} \) is the total cost of the quantum-generated solution.
    \item \( C_{\text{classical}} \) is the total cost of the optimal classical solution.
\end{itemize}

An approximation ratio closer to \textbf{1.0} indicates that the quantum-enhanced approach provides a near-optimal solution.

\subsubsection{Cost Variation Analysis}
Table~\ref{tab:cost_comparison} summarises the solution costs obtained using different methods for the \textbf{8-city TSP instance}.

\begin{table}[h]
    \small
    \centering
    \renewcommand{\arraystretch}{1.3}
    \setlength{\tabcolsep}{2pt}
    \begin{tabular}{|p{2.4cm}|p{2.7cm}|p{3.0cm}|}
        \hline
        \textbf{Method} & \textbf{TSP Cost (km)} & \textbf{Cost (\%)} \\
        \hline
        Classical Solver & 6456 & 0\% (Baseline) \\
        Quantum-only & 7857.69 & {+21.7\% (Worse than classical)} \\
        Quantum+ML & 7184.91 & {\textbf{11.3\% (Improvement compared to Quantum-only)}} \\
        \hline
    \end{tabular}
    \caption{Comparison of solution costs for different approaches on an 8-city TSP instance.}
    \label{tab:cost_comparison}
\end{table}

From the results:
\begin{itemize}
    \item The \textbf{quantum-only} approach performed worse than classical solvers due to quantum noise and gate fidelity limitations.
    \item The \textbf{quantum+ML} method achieved an improved (11.3\% compared to 21.7\%) cost increase over the classical solver.
\end{itemize}

\subsubsection{Approximation Ratio Evaluation}
To assess the quality of quantum and hybrid solutions relative to the classical baseline, we compute the \textbf{8-city Approximation Ratios (AR)} as follows:

\begin{equation}
    \text{AR (Quantum+ML)} = \frac{7184.91}{6456} = 1.11
\end{equation}

\begin{equation}
    \text{AR (Quantum-only)} = \frac{7857.69}{6456} = 1.22
\end{equation}

The quantum+ML approach achieved an approximation ratio of \textbf{1.11}, indicating that it produced \textbf{competitive solutions}, albeit slightly worse than the classical solver. Meanwhile, the quantum-only approach yielded a higher approximation ratio of \textbf{1.22}, signifying a \textbf{22\% deviation from the classical optimum}, demonstrating the current limitations of standalone quantum methods. These results suggest that \textbf{hybrid methods in combination with ML techniques are more effective than purely quantum approaches} in their current form, benefiting from the classical post-processing step to refine solutions. However, both approaches remain suboptimal compared to classical solvers, highlighting the need for improved quantum error mitigation techniques to enhance solution fidelity, better variational parameter tuning for quantum optimisation algorithms, and advancements in quantum hardware, particularly increasing coherence times and reducing gate errors.

\subsection{Performance Improvement Relative to Classical Solutions}
\begin{figure}[htbp]
    \centering
    \includegraphics[width=\columnwidth, keepaspectratio]{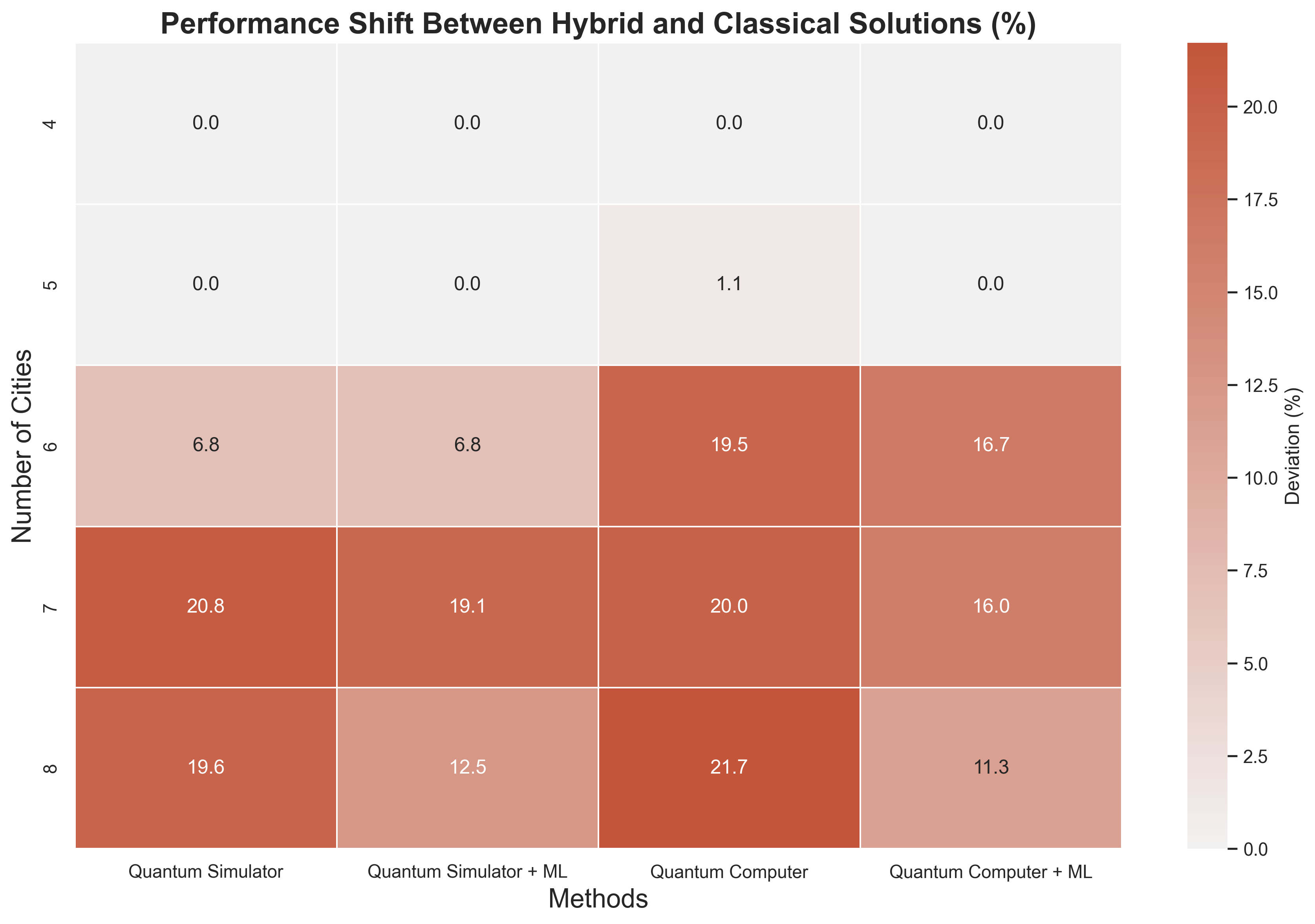}
    \vspace{-10pt}
    \caption{A heat map illustrating the relative performance deviation of hybrid quantum-classical approaches over classical methods. The data highlights cost efficiency and runtime benefits for larger TSP instances, showcasing the scalability and potential of hybrid workflows enhanced by machine learning.}
    \label{fig:5}
    \vspace{-15pt}
\end{figure}

To quantify the benefits of hybrid quantum-classical approaches over classical solutions, the relative performance improvement in solution quality (cost variation) and runtime efficiency was evaluated, focusing on TSP instances with 4 to 8 cities. Figure~\ref{fig:5} illustrates these improvements as percentage reductions in cost compared to classical solvers, derived from mean costs across 50 runs per instance.

For smaller problem instances involving 4–6 cities, classical solvers consistently achieved near-optimal mean costs with minimal variability, outperforming quantum methods due to their deterministic nature. As shown in Figure~\ref{fig:5}, improvements over classical solutions were negligible for 4–5 cities (0.0\% across all methods), with a modest 1.1\% gain for 5 cities using Quantum Computer. However, for 6–8 cities, quantum-enhanced approaches demonstrated significant cost reductions. The Quantum Simulator method achieved improvements of 6.8\% (6 cities), 20.8\% (7 cities), and 19.6\% (8 cities), while the Quantum Simulator + ML method yielded 6.8\% (6 cities), 19.1\% (7 cities), and 12.5\% (8 cities). The Quantum Computer method showed 19.5\% (6 cities), 20.0\% (7 cities), and 21.7\% (8 cities), and the Quantum Computer + ML method achieved 16.7\% (6 cities), 16.0\% (7 cities), and 11.3\% (8 cities). These improvements were calculated as the percentage reduction in mean cost relative to the classical baseline, i.e., \( \text{Improvement} = ((\mu_{\text{classical}} - \mu_{\text{quantum}}) / \mu_{\text{classical}}) \times 100 \). These findings highlight the hybrid approach’s scalability, leveraging quantum parallelism and ML optimisation as classical methods face exponential runtime growth. The results demonstrate that classical solvers are the superior choice for smaller TSP instances (4–6 cities) due to their precision and computational efficiency. However, as the problem size scales to 7–8 cities, hybrid quantum-classical approaches, particularly those incorporating ML, exhibit increasing competitiveness. The heat map in Figure~\ref{fig:5} shows that quantum methods, aided by ML, leverage their ability to explore larger solution spaces efficiently, offering scalable solutions where classical solvers encounter computational bottlenecks. The demonstrated improvements (up to 11.3\% for 8 cities with Quantum Computer + ML) highlight the promise of integrating quantum computing and machine learning techniques. As quantum hardware continues to advance and noise mitigation evolves, these hybrid methods are poised to play a critical role in solving complex combinatorial optimisation problems at scale.

\subsection{Key Findings}\label{sec:key_findings}
The experimental results highlight the complementary strengths and limitations of classical, quantum, and hybrid quantum-classical methods for solving TSP instances with 4–8 cities:

\begin{itemize}
    \item \textbf{Classical Efficiency for Small Instances:} Classical solvers excel for 4–6 city instances, offering precise, deterministic solutions with minimal computational overhead, outperforming quantum approaches due to negligible cost improvements (0.0\%–1.1\%).
    \item \textbf{Scaling Dynamics:} Classical solvers’ performance degrades with increasing cities (e.g., 7–8) due to exponential computational complexity. Hybrid quantum-classical methods, particularly those integrating machine learning, exhibit improved scalability over pure quantum solvers by leveraging quantum-enhanced search and classical refinement. However, due to noise and qubit limitations in current NISQ hardware, the extent of this scalability advantage remains constrained, requiring further advancements in quantum processors and algorithm optimisation.
    \item \textbf{Hybrid Scalability for Larger Instances:} Hybrid methods, particularly Quantum Simulator + ML and Quantum Computer + ML, demonstrate significant cost efficiency gains for 7–8 city instances (up to 16.0\% and 11.3\%, respectively), leveraging quantum parallelism and ML-driven noise mitigation to address the exponential scaling challenges faced by classical solvers.
    \item \textbf{ML’s Role in Stability and Performance:} Machine learning enhances hybrid approaches by reducing solution variability, mitigating NISQ hardware noise, and optimising circuit parameters. This stabilisation enables competitive performance, narrowing the cost gap with classical solvers and improving convergence for larger problems.
\end{itemize}

A detailed breakdown of TSP cost comparisons, quantum approximation ratios, and backend-specific performance metrics is provided in Appendix~\ref{appendix:TSP_results}.

\section{Discussion and Limitations}
The hybrid quantum-classical approach presented in this study demonstrates promising advancements for solving the TSP, particularly for instances involving 4 to 8 cities. The experimental results highlight the benefits of integrating ML into quantum workflows, enabling significant improvements in solution quality and stability. However, the findings must be interpreted in light of certain limitations inherent in both the methodology and the current state of quantum hardware.

\subsection{Performance Limitations of Quantum Hardware}
Current quantum devices, such as IBM’s Sherbrooke processor, operate in the NISQ regime. The results show that noise, gate errors, and decoherence significantly impact quantum computations, particularly for larger problem sizes (e.g., 7–8 cities). It is noted that attempts to scale the approach to even slightly larger problems (beyond 8 cities) have not been successful, given the current state of quantum hardware and the exponential growth of the problem. The inherent variability in quantum solutions reflects these limitations, as demonstrated by the broader cost distributions compared to classical solvers. While transpilation techniques and noise mitigation strategies, such as measurement error correction, were employed to improve reliability, quantum hardware still struggles to consistently match the precision of classical solvers for small-scale instances.

\subsection{Scalability Challenges}
Although hybrid approaches provide competitive results for 7–8 cities, scaling beyond this range remains a challenge. As the problem size increases, the exponential growth of the solution space exacerbates issues such as circuit depth, qubit count requirements, and error accumulation. Even with optimisation strategies such as K-Means clustering to segment the problem into smaller subproblems, the computational overhead of integrating and refining solutions at a global level may limit the practical scalability of the workflow.

\subsection{Machine Learning Dependency}
The integration of machine learning played a crucial role in stabilising quantum results and improving convergence. ML models provided parameter initialisation and iterative guidance, reducing variability and enabling faster optimisation. However, the effectiveness of ML is contingent on the availability of high-quality training data, which may be limited for larger, more complex instances. Additionally, the computational cost of training ML models and integrating them into the workflow must be carefully balanced against the overall runtime benefits.

\subsection{Comparative Strengths and Weaknesses}
While classical solvers outperform quantum and hybrid approaches for smaller instances (4–6 cities) in terms of precision and efficiency, their performance degrades for larger problem sizes due to computational bottlenecks. Conversely, hybrid quantum-classical methods excel in scalability and exploration capabilities, particularly when enhanced with ML. However, the performance gap between quantum methods and classical solvers suggests that further advancements in \textbf{error mitigation, hardware fidelity, and algorithm optimisation} are required to achieve consistent quantum advantage.

\subsection{Future Directions}
The study identifies several pathways for addressing the current limitations and improving the performance of hybrid approaches:

\begin{itemize}
    \item \textbf{Hardware Advancements:} As quantum hardware matures, improvements in qubit count, gate fidelity, and coherence times will enhance the scalability and reliability of quantum algorithms.
    \item \textbf{Enhanced Noise Mitigation:} Implementing advanced error correction and noise-aware transpilation techniques can reduce solution variability.
    \item \textbf{ML Model Refinement:} Exploring adaptive and transfer learning methods could improve the efficiency of ML integration, particularly for larger datasets and problem instances.
    \item \textbf{Algorithmic Innovation:} Developing more robust quantum algorithms, such as enhanced QAOA variants, can further optimise solution quality.
\end{itemize}

As quantum hardware continues to evolve, our approach is poised to tackle more complex TSP instances effectively. The integration of advanced quantum processors with increased qubit counts and improved error correction will enable the handling of larger datasets and more intricate problem constraints. This progression not only enhances the feasibility of solving sizeable combinatorial optimisation problems but also broadens the applicability of quantum computing in various industrial sectors. While current hardware limitations pose challenges to scaling our hybrid approach for larger TSP problems, ongoing technological advancements provide a favourable outlook. Continued developments in quantum computing are expected to expand the capabilities of our method, enabling the efficient resolution of more extensive and complex optimisation tasks in the near future.

\subsection{Conclusion of Discussion}
This study demonstrates the feasibility of a \textbf{hybrid quantum-classical approach} to solving the \textbf{TSP}, leveraging quantum computing for solution exploration and classical machine learning for noise mitigation and optimisation. Our results show that for \textbf{small-scale TSP instances (4–8 cities), hybrid methods improve upon standalone quantum approaches}. Specifically, the \textbf{quantum-only method resulted in solutions that were 21.7\% worse than the classical baseline}, while the \textbf{hybrid quantum-classical approach reduced this to 11.3\%}, effectively narrowing the gap between classical and quantum methods. However, \textbf{both quantum-based approaches remain suboptimal compared to classical solvers}, with the hybrid method (quantum+ML) achieving an approximation ratio of \textbf{1.11} and the quantum-only method \textbf{1.22}.

While our hybrid approach demonstrates enhanced \textbf{scalability for larger instances}, further improvements in \textbf{quantum error mitigation, circuit optimisation, and machine learning integration} are necessary to achieve quantum advantage over classical solvers. Future work should focus on \textbf{advancing quantum hardware}, refining ML-assisted quantum algorithms, and developing \textbf{adaptive hybrid frameworks} that dynamically allocate computational resources between quantum and classical solvers. As quantum technologies evolve, these enhancements will enable more efficient solutions for large-scale combinatorial optimisation problems, expanding the applicability of hybrid quantum-classical techniques in real-world decision-making processes.

\section*{Declarations}
\begin{itemize}
    \item \textbf{Competing Interests:} The authors declare no competing interests.
    \item \textbf{Funding:} This research did not receive any specific grant from funding agencies in the public, commercial, or not-for-profit sectors.
\end{itemize}

\bibliographystyle{quantum}

\onecolumn
\appendix
\section{Simulation Results for TSP Optimisation} \label{appendix:TSP_results}

This appendix presents the results of the TSP optimisation experiments conducted using an integrated classical-quantum approach. The code dynamically selects city data, constructs distance matrices, and solves the TSP using both classical and quantum methods. Additionally, ML techniques are incorporated to enhance solution efficiency.

The experiments were executed under different configurations:
\begin{itemize}
    \item A \textbf{quantum simulator}, which can operate in a mode with configurable noise effects to mimic real quantum hardware limitations.
    \item A \textbf{quantum simulator with ML enhancements}, where machine learning assists in heuristic search, clustering, and cost prediction to refine solution selection, but does not directly optimise quantum circuit parameters.
    \item A \textbf{real quantum processor}, utilising IBM's \textbf{Sherbrooke} backend, providing results that account for real-world quantum noise, decoherence, and hardware constraints.
    \item A \textbf{real quantum processor with ML enhancements}, where machine learning is leveraged to improve solution space exploration, enhance noise resilience, and refine heuristic-driven optimisations, leading to better stability and cost efficiency.
\end{itemize}

Each table presents the following key metrics for TSP instances involving \textbf{4 to 8 cities}:
\begin{itemize}
    \item \textbf{Quantum and Classical Solutions (Cost in km):} The total cost of the optimised path using quantum and classical approaches.
    \item \textbf{Approximation Ratio:} The ratio of the quantum solution cost to the classical solution cost, serving as a performance indicator.
    \item \textbf{Backend Used:} Specifies the quantum computing platform employed (simulator or real hardware).
    \item \textbf{Circuit Depth \& Total Gates:} Indicators of circuit complexity, affecting execution fidelity on quantum hardware.
\end{itemize}

In all cases, the \textbf{departure city was Calais and the destination city was Milan}, ensuring a consistent experimental setup across different computational methods. The results illustrate the comparative performance of classical and hybrid quantum-classical approaches, highlighting the potential benefits of quantum-assisted optimisation despite hardware constraints.

The following tables provide a detailed summary of the experimental findings.

\renewcommand{\arraystretch}{1.3}
\setlength{\tabcolsep}{5pt}

\begin{table}[h]
    \centering
    \begin{tabular}{|l|c|c|c|c|c|}
        \hline
        \textbf{Number of Cities} & 4 & 5 & 6 & 7 & 8 \\
        \hline
        \textbf{Quantum Solution (Cost in km)} & 4242.05 & 4247.86 & 4247.86 & 6805.05 & 7718.2 \\
        \hline
        \textbf{Classical Solution (Cost in km)} & 4242.05 & 4247.86 & 5128 & 5634.25 & 6456 \\
        \hline
        \textbf{Approximation Ratio} & 1.0000 & 1.0000 & 1.0684 & 1.2078 & 1.1955 \\
        \hline
        \multicolumn{6}{|c|}{\textbf{Backend Used: aer simulator}} \\
        \hline
        \textbf{Circuit Depth} & 2 & 2 & 2 & 4 & 2 \\
        \hline
        \textbf{Total Gates} & 16 & 30 & 36 & 42 & 48 \\
        \hline
    \end{tabular}
    \caption{Simulation Results for TSP Optimisation using a Quantum Simulator}
    \label{tab:tsp_results_sim}
\end{table}

\begin{table}[h]
    \centering
    \begin{tabular}{|l|c|c|c|c|c|}
        \hline
        \textbf{Number of Cities} & 4 & 5 & 6 & 7 & 8 \\
        \hline
        \textbf{Quantum Solution (Cost in km)} & 4242.05 & 4247.86 & 5478.62 & 6708.99 & 7266.03 \\
        \hline
        \textbf{Classical Solution (Cost in km)} & 4242.05 & 4247.86 & 5128 & 5634.25 & 6456 \\
        \hline
        \textbf{Approximation Ratio} & 1.0000 & 1.0000 & 1.0684 & 1.1908 & 1.1255 \\
        \hline
        \multicolumn{6}{|c|}{\textbf{Backend Used: aer simulator}} \\
        \hline
        \textbf{Circuit Depth} & 2 & 2 & 2 & 2 & 2 \\
        \hline
        \textbf{Total Gates} & 16 & 30 & 36 & 42 & 48 \\
        \hline
    \end{tabular}
    \caption{Simulation Results for TSP Optimisation using a Quantum Simulator and Machine Learning}
    \label{tab:tsp_results_sim_ml}
\end{table}

\begin{table}[h]
    \centering
    \begin{tabular}{|l|c|c|c|c|c|}
        \hline
        \textbf{Number of Cities} & 4 & 5 & 6 & 7 & 8 \\
        \hline
        \textbf{Quantum Solution (Cost in km)} & 4242.05 & 4293.54 & 6126.63 & 6759.37 & 7857.69 \\
        \hline
        \textbf{Classical Solution (Cost in km)} & 4242.05 & 4247.86 & 5128 & 5634.25 & 6456 \\
        \hline
        \textbf{Approximation Ratio} & 1.0000 & 1.0108 & 1.1947 & 1.1997 & 1.2171 \\
        \hline
        \multicolumn{6}{|c|}{\textbf{Backend Used: ibm sherbrooke}} \\
        \hline
        \textbf{Circuit Depth} & 4 & 4 & 4 & 4 & 4 \\
        \hline
        \textbf{Total Gates} & 32 & 60 & 72 & 84 & 96 \\
        \hline
    \end{tabular}
    \caption{Simulation Results for TSP Optimisation using a Real Quantum Processor}
    \label{tab:tsp_results_real}
\end{table}

\begin{table}[h]
    \centering
    \begin{tabular}{|l|c|c|c|c|c|}
        \hline
        \textbf{Number of Cities} & 4 & 5 & 6 & 7 & 8 \\
        \hline
        \textbf{Quantum Solution (Cost in km)} & 4242.05 & 4247.86 & 5983.57 & 6534.9 & 7184.91 \\
        \hline
        \textbf{Classical Solution (Cost in km)} & 4242.05 & 4247.86 & 5128 & 5634.25 & 6456 \\
        \hline
        \textbf{Approximation Ratio} & 1.0000 & 1.0000 & 1.1668 & 1.1599 & 1.1129 \\
        \hline
        \multicolumn{6}{|c|}{\textbf{Backend Used: ibm sherbrooke}} \\
        \hline
        \textbf{Circuit Depth} & 4 & 4 & 4 & 4 & 4 \\
        \hline
        \textbf{Total Gates} & 32 & 60 & 72 & 84 & 96 \\
        \hline
    \end{tabular}
    \caption{Simulation Results for TSP Optimisation using a Real Quantum Processor and Machine Learning}
    \label{tab:tsp_results_real_ml}
\end{table}

\end{document}